\documentclass[useAMS,usenatbib]{mn2e}
\usepackage[a4paper,totalwidth=480pt,totalheight=680pt]{geometry}
\usepackage{amssymb}
\usepackage{bbm}
\usepackage{txfonts}
\usepackage{graphicx}
\usepackage{natbib}
\usepackage{if then}
\usepackage{longtable}
\usepackage{color}
\usepackage{hyperref}
\usepackage{ulem}
\usepackage{verbatim}
\usepackage{bm}
\usepackage{lastpage}
\usepackage{algorithm,algorithmicx,algpseudocode}

\def\del#1{{}}
\def\mnras{MNRAS}
\newcommand{\apj}{ApJ}

\newcommand{\aap}{A{\&}A}

\newcommand{\prd}{Phys. Rev. D}
\newcommand{\apjl}{ApJL}
\newcommand{\apjs}{ApJS}

\newcommand{\vect}[1]{{\bm{#1} }}
\newcommand{\mat}[1]{{\mathbf{#1}}}

\title{Matrix-free Large Scale Bayesian inference in cosmology}
\author[Jens~Jasche and
	Guilhem~Lavaux]
       {Jens~Jasche$^{1,2}$ and
       Guilhem~Lavaux$^{1,2}$ \\
   \\
$^{1}$ CNRS, UMR 7095, Institut d'Astrophysique de Paris, F-75014, Paris, France\\
$^{2}$ Sorbonne Universit\'es, UPMC Univ Paris 6, UMR 7095, IAP, F-75014, Paris, France\\
}
\begin{document}
\date{Accepted 20?? December ??.
      Accepted 20?? December ??;
      in original form 20?? October ??}

\pagerange{\pageref{firstpage}--\pageref{LastPage}} \pubyear{2014}

\maketitle

\label{firstpage}

\begin{abstract}
In this work we propose a new matrix-free implementation of the Wiener sampler which is traditionally applied to high dimensional analysis when signal covariances are unknown. Specifically, the proposed method addresses the problem of jointly inferring a high dimensional signal and its corresponding covariance matrix from a set of observations. Our method implements a Gibbs sampling adaptation of the previously presented messenger approach, permitting to cast the complex multivariate inference problem into a sequence of uni-variate random processes. In this fashion, the traditional requirement of inverting high dimensional matrices is completely eliminated from the inference process, resulting in an efficient algorithm that is trivial to implement.  
Using cosmic large scale structure data as a showcase, we demonstrate the capabilities of our Gibbs sampling approach by performing a joint analysis of three dimensional density fields and corresponding power-spectra from Gaussian mock catalogues. These tests clearly demonstrate the ability of the algorithm to accurately provide measurements of the three dimensional density field and its power-spectrum and corresponding uncertainty quantification. Moreover, these tests reveal excellent numerical and statistical efficiency which will generally render the proposed algorithm a valuable addition to the toolbox of large scale Bayesian inference in cosmology and astrophysics.
\end{abstract}


\section{Introduction}
Ever increasing amounts and precision of modern cosmological and astrophysical data
demands fast and robust methods to address corresponding large scale inference problems of analysing these 
observations and extracting new knowledge on our Universe.  
Particularly, the Wiener filter has become a standard tool for the analysis of large data sets, often involving many millions of parameters, with widespread applications in cosmology and astrophysics.
Even though relying on a linear data model and Gaussian statistics, the Wiener filter approach is still
a standard and well valued method when requiring a robust approach for the inference of high dimensional signals as occurring in the analysis of large scale structure (LSS) or cosmic microwave background (CMB) data.
For this reason, Wiener filtering has been frequently applied to a variety of large scale structure analysis problems, specifically the inference of the three dimensional density field from galaxy observations \citep[see e.g.][]{1994ApJ...423L..93L,WienerFSL,1995MNRAS.272..885F,1993ApJ...415L...5G,1995ApJ...449..446Z,1995MNRAS.272..885F,HOFFMAN1994,1995MNRAS.277..933S,1999ApJ...520..413Z,2002MNRAS.331..901Z,2004MNRAS.352..939E,KITAURA2008,2006MNRAS.373...45E,KITAURA2009,JASCHE2010PSPEC,JASCHE_SPEC2013}.

In the field of CMB analysis, the Wiener filter is frequently employed as a map making algorithm or for the joint inference of temperature fluctuations and corresponding power-spectra \citep[see e.g.][]{2004ApJS..155..227E,JEWELL2004,ODWYER2004,WANDELT2004,2004ApJS..155..227E,SMITH2007,LARSON2007,ERIKSEN2007,ELSNER2013}.
Similar approaches have also been used for the optimal reconstruction of images from radio Interferometry \citet{2006ApJS..162..401S,SUTTER2013} or to generate improved maps of the galactic Faraday emission \citet{2012A&A...542A..93O}.
Furthermore, the Wiener filter also constitutes an integral part of the recently presented general purpose statistical analysis framework NIFTY (Numerical Information Field Theory) \citep[][]{SELIG2013}.  

Traditionally, numerical implementations of the Wiener filter rely on Krylov space methods, such as conjugate gradients, to invert matrices and solve high dimensional systems of linear equations \citep[see e.g.][and references therein]{KITAURA2008}. 
However, implementation and testing of these numerically demanding methods also constitutes a certain hurdle and generally requires some investment into code development before such methods can be used for specific scientific applications.

Recently a particular elegant and simple way of implementing the Wiener filter via a messenger method has been proposed, which remedies the hurdle of implementing numerical matrix inversion techniques for very high dimensional systems \citep[][]{ELSNER2013}. In particular, \citet{ELSNER2013} proposes to introduce a messenger field to mediate between different preferred orthogonal bases, in which signal and noise covariance matrices can be expressed conveniently. In this approach information from the data is transmitted to the signal via a messenger field, that can generally be transformed efficiently from one bases representation to another. In this fashion the algorithm avoids the requirement to apply the inverse Wiener covariance matrix to data.

In this work, we will pick up these ideas and propose an efficient and easy to implement Gibbs sampling approach to address Bayesian large scale inference problems in cosmology or astrophysics.  Our primary intention is to propose a simple, nevertheless powerful, algorithm for the joint inference of a signal and corresponding covariance matrix from observations that can be implemented and operated by everyone, even inexperienced users. Specifically, in this work, we will exemplify the performance of this algorithm in case of a LSS analysis aiming at the joint inference of the three dimensional density field and cosmological power-spectrum from galaxy surveys.

The introduction of a messenger field yields an augmented Wiener posterior distribution, whose structure lends itself to a ideal multiple block sampling approach. In a first step the messenger field is realised from a normal distribution conditional on the observation in real space, followed by a second step of sampling the signal conditional on the previously sampled messenger field in Fourier-space. Here, the augmented Wiener posterior distribution is chosen such, that sampling these two random fields can be trivially achieved by generating a sequence of uni-variate normal random variates in their respective basis representations. Iterating these processes will then provide samples from the Wiener posterior without the need of performing matrix inversions or any other multi-parameter operation, except for basis transformations. Further we will complement this algorithm, by a power-spectrum sampling method to jointly infer the signal and its covariance matrix. Likewise, as described in the following, this 
algorithm will also only require the ability to generate uni-variate normal and inverse gamma variates.

Consequently, we arrive at an efficient algorithm, which, at every stage, reduces the full joint problem to a sequence of independent uni-variate sub-problems. The advantage of this algorithm lies in its ease of implementation, thus greatly reducing the required investment in code development for large scale Bayesian inference projects. 
In the following we will describe the implementation of this algorithm and exemplify it in case of a mock LSS analysis.
The paper is structured as follows.
In section \ref{augmented_posterior} we will describe the messenger approach and the resulting augmented Wiener posterior distribution.
Section \ref{implementation}, describes the numerical implementation of the proposed method.
To estimate the performance of the algorithm in a realistic scenario, as an example, we will apply it to an artificial
galaxy survey, which will be described in section \ref{mock_Data}. In the following section \ref{Results}, we will discuss the results of these tests and conclude the paper with a summary and conclusion in section \ref{conclusion}.

\section{The augmented Wiener posterior}
\label{augmented_posterior}
As described in the introduction, the aim of this work is to present an easy to implement algorithm for the large scale Bayesian problem of jointly inferring a signal and its covariance matrix in a high dimensional setting. Specifically,
we aim at exploring the joint posterior distribution \(\Pi\left(\vect{s},S|\vect{d}\right)\) of the signal \(\vect{s}\) and its covariance matrix \(S\) conditional on observations \(\vect{d}\). Using Bayes rule this posterior distribution can be rewritten as:
\begin{eqnarray}
\Pi\left(\vect{s},\mat{S}|\vect{d}\right) =  \Pi\left(\mat{S}\right)\,\Pi\left(\vect{s}|\mat{S} \right)\, \frac{\Pi\left(\vect{d}|\vect{s}\right)}{\Pi\left(\vect{d} \right)} \, ,
\label{eq:joint_posterior}
\end{eqnarray}
where \(\Pi\left(\mat{S}\right)\) is the signal covariance prior, \(\Pi\left(\vect{s}|\mat{S} \right)\) is the signal prior and \(\Pi\left(\vect{d}|\vect{s}\right)\) is the likelihood normalised by the evidence \(\Pi\left(\vect{d} \right)\). Note, that observations \(\vect{d}\) are assumed to be conditionally independent of the signal covariance matrix once the signal is given, specifically \(\Pi\left(\vect{d}|\mat{S},\vect{s}\right)=\Pi\left(\vect{d}|\vect{s}\right)\).
In the following we will assume linear data models of the form:
\begin{equation}
\vect{d}=\mat{R}\,\vect{s} + \vect{\epsilon} \,
\end{equation}
where \(\mat{R}\) is a linear measurement response operator and \(\epsilon\) is a normally distributed noise vector with zero mean and noise covariance matrix \(N\). Further, assuming a Gaussian prior for the signal yields the famous Wiener posterior for the inference of the signal \(\vect{s}\) given as:
\begin{equation}
\Pi\left(\vect{s}|\mat{S} \right)\, \frac{\Pi\left(\vect{d}|\vect{s}\right)}{\Pi\left(\vect{d} \right)} = \frac{\mathrm{e}^{-\frac{1}{2} \vect{s}^T \mat{S}^{-1}\vect{s}}}{\sqrt{\mathrm{det}2\pi\,\mat{S}}}\, \frac{\mathrm{e}^{-\frac{1}{2} \left(\vect{d} - \mat{R}\vect{s}\right)^T\mat{N}^{-1}\left(\vect{d} - \mat{R}\vect{s}\right)}}{\sqrt{\mathrm{det}2\pi\,\mat{N}}}\, = \Pi\left(\vect{s}|\mat{S},\vect{d} \right).
\label{eq:target_wiener_post}
\end{equation}
Complication in inferring signals from the Wiener posterior arises for the analysis of large and complex data sets  
since the sizes of signal and noise covariance matrices scale quadratically with the number of signal parameter and data points \citep{ELSNER2013}. This fact generally renders storage and processing of dense systems impractical.
Although it is often possible to find a set of bases in which the respective noise and signal covariances can be represented by sparse matrices, it is generally not possible to jointly represent them as sparse systems in  a single basis. As a consequence traditional MCMC methods rely on the implementation of complex numerical algorithms such as Krylov space methods or gradient based hybrid Monte Carlo approaches to solve the corresponding sets of linear equations \citep[for examples in cosmological applications see e.g.][]{2004ApJS..155..227E,JEWELL2004,ODWYER2004,WANDELT2004,2004ApJS..155..227E,LARSON2007,ERIKSEN2007,ELSNER2013,1994ApJ...423L..93L,WienerFSL,1995MNRAS.272..885F,1993ApJ...415L...5G,1995ApJ...449..446Z,1995MNRAS.272..885F,HOFFMAN1994,1995MNRAS.277..933S,1999ApJ...520..413Z,2002MNRAS.331..901Z,2004MNRAS.352..939E,KITAURA2008,2006MNRAS.373...45E,KITAURA2009,JASCHE2010PSPEC,JASCHE_SPEC2013}. 

In this situation \citet{ELSNER2013} proposed to introduce a normally distributed messenger field \(\vect{t}\) with covariance matrix \(\mat{T}\), to mediate between the respective bases in which \(\mat{S}\) and \(\mat{N}\) can be represented as sparse systems. In particular, the covariance matrix \(\mat{T}\) is chosen to be proportional to the diagonal matrix, a property which is conserved under orthogonal basis transforms.	
The introduction of this additional random field to the inference process yields an augmented Wiener posterior for the joint inference of the signal \(\vect{s}\) and the messenger field \(\vect{t}\) given as:
\begin{equation}
\Pi\left(\vect{s},\vect{t}|\mat{S},\mat{T},\vect{d} \right) = \frac{\mathrm{e}^{-\frac{1}{2} \vect{s}^T\mat{S}^{-1}\vect{s}}}{\sqrt{\mathrm{det}2\pi\,\mat{S}}}\,\frac{\mathrm{e}^{-\frac{1}{2} \left(\vect{s}-\vect{t}\right)^T\mat{T}^{-1}\left(\vect{s}-\vect{t}\right)}}{\sqrt{\mathrm{det}2\pi\,\mat{T}}}\, \frac{\mathrm{e}^{-\frac{1}{2} \left(\vect{d} - \mat{R}\vect{t}\right)^T\mat{\tilde{N}}^{-1}\left(\vect{d} - \mat{R}\vect{t}\right)}}{\sqrt{\mathrm{det}2\pi\,\mat{\tilde{N}}}}\, . 
\label{eq:augmented_wiener_post}
\end{equation} 	 
Note, if the messenger covariance matrix $\mat{T}$ is proportional to a diagonal matrix \(\mat{T}=\tau\,\mat{1}\), marginalising over the messenger field will yield the target distribution, given in equation (\ref{eq:target_wiener_post}), if the augmented noise covariance \(\mat{\tilde{N}}\) is chosen as:
\begin{equation}
\mat{\tilde{N}}=\mat{N} - \mat{R}^T \mat{T} \mat{R} \, .
\end{equation}
Furthermore, requiring the augmented noise covariance matrix \(\mat{\tilde{N}}\) to be positive definite,
yields:
\begin{equation}
0<\tau \leq \left[\left(\mat{R}^{-1}\right)^{T} \mat{N}\, \mat{R}^{-1}\right]_i \, \forall \, i \, ,
\end{equation}
specifically we choose \(\tau\) to be the minimum of all entries in \(\left(\mat{R}^{-1}\right)^{T} \mat{N}\, \mat{R}^{-1}\) in the observed domain.

\section{A large scale Gibbs sampler}
\label{implementation}
This section describes the derivation of our algorithm and describes its numerical implementation.

\subsection{Generating signal realisations}
The augmented Wiener posterior distribution given in equation (\ref{eq:augmented_wiener_post}) lends itself to a  multiple block sampling approach. In particular, the problem of jointly exploring the augmented Wiener posterior can be reduced to the task of sequentially sampling the signal field \(\vect{s}\) and the messenger field \(\vect{t}\). Specifically we propose to generate random variates of the respective fields via the following two step sampling approach:
\begin{eqnarray}
\vect{s} &\curvearrowleft & \Pi\left(\vect{s}|\mat{S},\mat{T},\vect{t}, \vect{d} \right)=\Pi\left(\vect{s}|\mat{S},\mat{T},\vect{t} \right)\\
\vect{t} &\curvearrowleft & \Pi\left(\vect{t}|\mat{S},\mat{T},\vect{d},\vect{s} \right)=\Pi\left(\vect{t}|\mat{T},\vect{d},\vect{s} \right)\, .\
\label{eq:wiener_sampling_process}
\end{eqnarray} 
Iterating these processes will yield samples from the joint augmented Wiener posterior distribution. Marginalisation is then trivially achieved by simply discarding the respective realisations of the messenger field \(\vect{t}\), yielding  signal realisations \(\vect{s}\) correctly drawn from the target Wiener posterior given in equation (\ref{eq:target_wiener_post}). 
The important point to remark, as demonstrated by the augmented Wiener posterior distribution, given in equation (\ref{eq:augmented_wiener_post}), and as manifested by the proposed sampling procedure given in equation (\ref{eq:wiener_sampling_process}), is that information between data \(\vect{d}\) and signal \(\vect{s}\) is not transmitted directly between those two fields but is mediated via a third messenger field. As the messenger covariance matrix \(\mat{T}\) is diagonal in the respective bases in which signal and noise covariances can be described as sparse diagonal systems, random variates for signal and messenger fields can be generated by independently drawing uni-variate normal realisations for the individual elements of the respective fields in the respective bases. Specifically signal realisations are generated via the process:
\begin{eqnarray}
\hat{s}_i &\curvearrowleft &  \frac{\mathrm{e}^{-\frac{1}{2}\frac{\left(\hat{s}_i-\mu^{\hat{s}}_i\right)^2}{\left(\sigma^{\hat{s}}_i\right)^2}}}{\sqrt{2\pi\,\left(\sigma^{\hat{s}}_i\right)^2}}  \, \, \forall \, i \in M
\end{eqnarray}
with \(\mu^{\hat{s}}_i=\hat{S}_i/\left(\hat{S}_i + \hat{T}_i\right) \, \hat{t}_i\) and \(\left(\sigma^{\hat{s}}_i\right)^2=\hat{S}_i\,\hat{T}_i/\left(\hat{S}_i + \hat{T}_i\right)\). The index \(i\) labels the different elements of the respective vectors, and the hat operator indicates that all quantities have been transformed to the basis in which the Signal covariance matrix \(\mat{S}\) assumes its diagonal shape \(\mat{\hat{S}}\). 
In a analogous fashion uni-variate normal variates can be generated for the individual elements of the messenger field \(\vect{t}\) as:
\begin{eqnarray}
t_i &\curvearrowleft &  \frac{\mathrm{e}^{-\frac{1}{2}\frac{\left(t_i-\mu^{t}_i\right)^2}{\left(\sigma^{t}_i\right)^2}}}{\sqrt{2\pi\,\left(\sigma^{t}_i\right)^2}}  \, \, \forall \, i \in M
\end{eqnarray}
with:
\begin{eqnarray}
\mu^{t}_i=\left\{\begin{array}{cl} \frac{T_i}{\left(T_i\,R^2_i + \tilde{N}_i\right)} \, R_i\,d_i + \frac{\tilde{N}_i}{\left(T_i\,R^2_i + \tilde{N}_i\right)} \,s_i, & \mbox{if } R^2_i > 0\\ s_i, & \mbox{otherwise} \end{array}\right. 
\end{eqnarray}
and
\begin{eqnarray}
\left(\sigma^{t}_i\right)^2=\left\{\begin{array}{cl} \frac{T_i\,\tilde{N}_i}{\left(T_i\,R^2_i + \tilde{N}_i\right)}, & \mbox{if } R^2_i > 0\\ T_i, & \mbox{otherwise} \end{array}\right. 
\end{eqnarray}
Note, that for sampling the messenger field \(\vect{t}\) all quantities are given in the basis in which the noise covariance \(N\) becomes a diagonal matrix. It should also be remarked that the messenger covariances \(T_i\) and \(\hat{T}_i\) are the same only for normalised orthogonal transforms, otherwise they differ by the multiplicative normalisation constant.

Consequently, generating random signal variates \(\vect{s}\) from the Wiener posterior given in equation (\ref{eq:target_wiener_post}) only relies on the ability to draw uni-variate Gaussian random numbers and to perform
 orthonormal basis transformations to switch between different basis representations. In typical cosmological applications these orthonormal basis transformations are used to switch between real and Fourier space, which is achieved via fast and efficient implementations of the fast Fourier or Spherical Harmonic transformation algorithms \citep[see e.g.][]{ODWYER2004,JEWELL2009,KITAURA2008,JASCHE_SPEC2013}. 

 As can be seen from the derivation presented above, at no point does our approach rely on the storage and inversion of covariance matrices. A pseudo code for the proposed signal sampling algorithm is given in Algorithm \ref{alg_signal_sampler}.

\begin{algorithm}
\caption{signal sampler}\label{alg_signal_sampler}
\begin{algorithmic}[1]
\Procedure{signal\_sampler}{$s,t$}%\Comment{}
   \For{$i = 0 \to  (M-1)$}
			\State $t_i = \mu^t_i +\sqrt{\left(\sigma_i^t\right)^2}\, G_i(0,1) $   \Comment{G(0,1) is a unit normal random number}
			 \EndFor\label{forloop_messenger1}
   
   \State $\hat{t}=\mathrm{ONT}(t)$ \Comment{ONT = Ortho-Normal-Transform}
   \State $\hat{s}=\mathrm{ONT}(s)$
   
  \For{$i = 0 \to  N$}%\Comment{We have the answer if r is 0}
      \State $\hat{s}_i = \mu^{\hat{s}}_i +\sqrt{\left(\sigma_i^{\hat{s}}\right)^2}\, G_i(0,1) $
   \EndFor\label{forloop_signal}

   \State $s=\mathrm{ONT^{-1}}(\hat{s})$
   \State \textbf{return} $s,t$ %\Comment{}
\EndProcedure
\end{algorithmic}
\end{algorithm}

\subsection{Sampling the signal covariance}
Once a realisation of the signal \(\vect{s}\) as generated by Algorithm \ref{alg_signal_sampler} is available, sampling the signal covariance matrix becomes a particularly trivial task in the basis where it assumes its diagonal form. As can be seen from the full joint posterior distribution given in equation (\ref{eq:joint_posterior}), the conditional signal covariance posterior solely depends on the covariance and signal prior once a signal realisation has been specified:
\begin{eqnarray} 
\Pi\left(\mat{S}|\vect{d},\vect{s}\right) = \Pi\left(\mat{S}|\vect{s}\right) \propto \Pi\left(\mat{S}\right)\,\Pi\left(\vect{s}|\mat{S} \right) \propto \Pi\left(\mat{\hat{S}}\right)\,\Pi\left(\vect{\hat{s}}|\mat{\hat{S}} \right)  \, ,
\label{eq:cond_cov_post}
\end{eqnarray}
where for the last proportionality we used the fact that the determinant of the Jacobian of the coordinate transform induced by the orthonormal transformation is one. For the analysis of the individual elements of the diagonal signal covariance matrix \(\hat{S}_i\) we propose to use Jeffreys' prior given by \( \Pi\left(\mat{\hat{S}}\right) = \prod_{i=0}^{M-1} \left( \hat{S}_i\right)^{-1}  \), which factorises in the individual matrix elements. Jeffreys' prior is a solution to a measure invariant scale transformation, and hence is a scale independent prior, as different scales have the same probability \citep[][]{JEFFREYS1946}. For this reason, Jeffreys prior constitutes a optimal choice for many applications, such as the inference of cosmological power-spectra, which constitute scale measurements, since it does not introduce any bias on a logarithmic scale \citep[also see discussions in][]{JASCHE2010PSPEC,JASCHE_SPEC2013}.

Since due to the diagonal shape of \(\mat{\hat{S}}\) in the corresponding basis  representation, also the second factor \(\Pi\left(\vect{\hat{s}}|\mat{\hat{S}} \right) \) in equation (\ref{eq:cond_cov_post}) factorises in the matrix elements \(\hat{S}_i\), all these elements can be sampled independently. In particular, simple algebraic manipulation of equation (\ref{eq:cond_cov_post}) reveals that the individual matrix elements \(\hat{S}_i\) have to be drawn from an inverse gamma distribution:   
\begin{eqnarray}
\hat{S}_i &\curvearrowleft & \frac{\left(\frac{1}{2}\hat{s}_i^2\right)^{\frac{1}{2}}}{\Gamma\left(\frac{1}{2}\right)}\left( \hat{S}_i\right)^{-\frac{3}{2}}  \mathrm{e}^{-\frac{1}{2}\frac{\hat{s}_i^2}{\hat{S}_i}}  \, \, \forall \, i \in M \, .
\end{eqnarray}
Again the complex joint sampling process of all signal covariance matrix elements can be reduced to the trivial task of independently realising inverse gamma variates. In particular, introducing the coordinate transformation \(\hat{u}_i=\hat{s}^2_i/\hat{S}_i\) yields a chi-square distribution which gives rise to the sampling algorithm outlined in Algorithm \ref{alg_signal_cov_sampler}. In particular, sequential iteration of algorithms \ref{alg_signal_sampler} and \ref{alg_signal_cov_sampler} yields samples of the joint posterior distribution of the signal and its covariance matrix conditional on data.
It should be remarked, that in some cases additional symmetries can be exploited to further reduce the required number of parameters to describe the signal covariance matrix. In particular, for cosmological applications one can exploit the homogeneity and isotropy of the Universe to average the covariance matrix over spherical shells in Fourier space. For a discussion of the inverse Gamma sampler in a cosmological setting and the required minor modifications to Algorithm \ref{alg_signal_cov_sampler} the reader is referred to \citep[see e.g.][]{JASCHE2010PSPEC,JASCHE_SPEC2013}

\begin{algorithm}
\caption{signal covariance sampler}\label{alg_signal_cov_sampler}
\begin{algorithmic}[1]
\Procedure{signal\_covariance\_sampler}{$s$}%\Comment{}
\State $\hat{s}=\mathrm{ONT}(s)$
   \For{$i = 0 \to  (M-1)$}
   			\State $\hat{u}_i=\left(G_i(0,1)\right)^2$
			\State $\hat{S}_i = \frac{\left|\hat{s}\right|^2_i}{\hat{u}_i} $
			 \EndFor\label{forloop_messenger2}
   
   \State \textbf{return} $\mat{\hat{S}}$ %\Comment{}
\EndProcedure
\end{algorithmic}
\end{algorithm}

\subsection{Improving statistical efficiency}
The algorithms, as outlined above, already provide a correct Markov Chain Monte Carlo approach to explore the joint 
distribution of a signal and corresponding signal covariance.
While this approach provably converges to the target posterior at all regimes probed by the data, it may take prohibitive computational time to generate a sufficient amount of independent samples in the low signal to noise regime \citep[for a discussion of this issue see e.g.][]{JEWELL2009,JASCHE_SPEC2013}. 
In particular, the variations in subsequent samples of the signal covariance are solely determined by signal variance, whereas the full joint posterior distribution is governed by signal variance and noise. 
As a consequence the algorithms described above permit rapid exploration of parameters in the high signal to noise regime, but yield poor statistical efficiency in low signal to noise regimes, where signal variance is typically less than noise. Typically this results in a prohibitively long correlation length of the sequence of sampled signal covariances in the low-signal to noise regime, requiring unfeasible long Markov Chains to generate sufficient numbers of independent samples \citep{JEWELL2009,JASCHE_SPEC2013}.
Fortunately, the messenger approach permits to devise a particularly simple approach to overcome these limitations via a simple change of coordinates. Rather than separating the steps of sampling the signal and covariance matrix, as was described above, here we propose to combine the sampling steps of the signal $\hat{\vect{s}}$ and its covariance matrix  $\mat{\hat{S}}$ conditional on a realisation of the messenger field $\hat{\vect{t}}$ by exploring the conditional posterior distribution:
\begin{eqnarray}
\Pi\left(\vect{s},\mat{S}|\vect{d},\vect{t}\right) &=& \Pi\left(\vect{s},\mat{S}|\vect{t}\right) \propto \Pi\left(\mat{\hat{S}}\right)\,\Pi\left(\hat{\vect{s}}|\mat{\hat{S}}\right)\,\Pi\left(\hat{\vect{t}}|\hat{\vect{s}}\right) \nonumber \\
&\propto& \prod_i \frac{1}{\left(\hat{S}_i\right)^{\frac{3}{2}}}\, \mathrm{e}^{-\frac{1}{2}\frac{\hat{s}_i^2}{\hat{S}_i}}\, \mathrm{e}^{-\frac{1}{2}\frac{\left( \hat{s}_i - \hat{t}_i \right)^2}{\hat{T}_i}}\, .
\label{eq:cond_messenger_dist_base}
\end{eqnarray}
Introducing the following change of coordinates \(\hat{s}_i=\sqrt{\hat{S}_i}\, \hat{x}_i \) then yields the transformed distribution:
\begin{eqnarray}
\Pi\left(\hat{\vect{x}},\mat{\hat{S}}|\hat{\vect{t}}\right) &\propto& \prod_i \sqrt{\hat{S}_i}\, \mathrm{e}^{-\frac{1}{2}\,\hat{x}_i^2}\, \mathrm{e}^{-\frac{1}{2}\frac{\left( \sqrt{\hat{S}_i}\, \hat{x}_i - \hat{t}_i \right)^2}{\hat{T}_i}}\, , 
\label{eq:cond_messenger_dist}
\end{eqnarray}
which again factorises in the individual elements. Exploring the joint distribution of \(\hat{\vect{x}}\) and \(\mat{\hat{S}}\)
can then again be achieved by sampling individual elements via a block sampling algorithm.
In the first step realisations for the \(\hat{\vect{x}_i}\) components are drawn via the following process:  
\begin{eqnarray}
\hat{x}_i &\curvearrowleft &  \frac{\mathrm{e}^{-\frac{1}{2}\frac{\left(\hat{x}_i-\mu^{\hat{x}}_i\right)^2}{\left(\sigma^{\hat{x}}_i\right)^2}}}{\sqrt{2\pi\,\left(\sigma^{\hat{x}}_i\right)^2}}  \, \, \forall \, i \in M
\end{eqnarray}
with:
\begin{eqnarray}
\mu^{\hat{x}}_i= \frac{\hat{T}_i}{\hat{S}_i+\hat{T}_i} \, \sqrt{\hat{S}_i} \hat{t}_i \, ,
\end{eqnarray}
and:
\begin{eqnarray}
\left(\sigma^{\hat{x}}_i\right)^2= \frac{\hat{T}_i}{\hat{S}_i+\hat{T}_i}  \, .
\end{eqnarray}
Conditional on these realisations of \(\hat{\vect{x}_i}\) samples for the elements \(\hat{S}_i\) of the signal covariance can be generated by drawing random variates from the conditional distribution:
\begin{eqnarray}
\Pi\left(\hat{S}_i|\hat{\vect{t}}_i , \hat{\vect{x}}_i\right) &\propto& \sqrt{\hat{S}_i} \mathrm{e}^{-\frac{1}{2}\frac{\left( \sqrt{\hat{S}_i}\, \hat{x}_i - \hat{t}_i \right)^2}{\hat{T}_i}}\, . 
\label{eq:cond_cov_dist_re_base}
\end{eqnarray}
Unfortunately, sampling this distribution directly is not possible. Consequently, the proposed Markov Chain Monte Carlo approach has to rely on a Metropolis-Hastings acceptance step.
For this purpose we introduce the change of coordinate \(\hat{u}_i=\sqrt{\hat{S}_i}\), which yields the distribution:
\begin{eqnarray}
\Pi\left(\hat{u}_i|\hat{\vect{t}}_i , \hat{\vect{x}}_i\right) &\propto& \left(\hat{u}_i\right)^2\,  \mathrm{e}^{-\frac{1}{2}\frac{\left( \hat{u}_i\, \hat{x}_i - \hat{t}_i \right)^2}{\hat{T}_i}} \propto \left(\hat{u}_i\right)^2\,  \mathrm{e}^{-\frac{1}{2}\frac{\left( \hat{u}_i\, - \frac{\hat{t}_i}{\hat{x}_i}\right)^2}{\hat{T}_i/\hat{x}^2_i}} \, . 
\label{eq:cond_cov_dist_re}
\end{eqnarray}
It can be seen, that the resultant distribution for \(\hat{u}_i\) is essentially a normal distribution multiplied by the factor \(\hat{u}^2_i\) which ensures that samples of the covariance matrix will be strictly positive definite. Although direct sampling from this distribution is not possible, our tests have shown, that generating proposals from a truncated normal distribution yields nearly ideal acceptance rates in a Metropolis-Hastings step. In particular we propose to use a independence sampler by generating proposals \(\hat{u}'_i\) via the process:
\begin{eqnarray}
\hat{u}'_i &\curvearrowleft & \Theta(\hat{u}'_i)\,  \mathrm{e}^{-\frac{1}{2}\frac{\left( \hat{u}_i\, - \frac{\hat{t}_i}{\hat{x}_i}\right)^2}{\hat{T}_i/\hat{x}^2_i}} \, ,
\end{eqnarray}
where \(\Theta(x)\) denotes the Heaviside function. Using this particular proposal distribution then yields the standard Metropolis-Hastings acceptance probability for each individual element:
\begin{eqnarray}
\alpha = \mathrm{min}\left(1, \left(\frac{\hat{u}'_i}{\hat{u}_i}\right)^2\right) \, .
\end{eqnarray}
As in the previous sections, the proposed algorithm only relies on independent updates in the Markov Chain and hence is trivial to implement. The pseudo code for this algorithm is given in algorithm \ref{alg_mixing_sampler}.
The proposed algorithm is optimal to sample the low signal to noise regime. In particular, introducing the initial change of coordinates \(\hat{s}_i=\sqrt{\hat{S}_i}\, \hat{x}_i \) moved the covariance matrix from the signal prior to the messenger posterior distribution in equation (\ref{eq:cond_messenger_dist}). As a consequence,  step size between subsequent covariance matrix samples is not determined by the prior variance but by the larger noise variance represented by \(\hat{T}_i\).

\begin{algorithm}
\caption{mixing sampler}\label{alg_mixing_sampler}
\begin{algorithmic}[1]
\Procedure{mixing\_sampler}{$t,\mat{\hat{S}}$}%\Comment{}
\State $\hat{t}=\mathrm{ONT}(t)$
   \For{$i = 0 \to  (M-1)$}
   			\State $\hat{x}_i=\mu^{\hat{x}}_i +\sqrt{\left(\sigma_i^{\hat{x}}\right)^2}\, G_i(0,1)$
   			\While{$\hat{u}'_i < 0 $}
    		 \State $\hat{u}'_i = \frac{\hat{t}_i}{\hat{x}_i} +\sqrt{\frac{\hat{T}_i}{\hat{x}^2_i}} \, G_i(0,1) $
  			\EndWhile
			\State $\hat{u}_i = \sqrt{\hat{S}_i}  $
			\State $\alpha_i = U_i(0,1)  $\Comment{ $U(0,1)$ is a unit random number}
			\If{$\left(\frac{\hat{u}'_i}{\hat{u}_i} \right)^2>\alpha_i$}
			 \State $ \hat{S}_i = \left(\hat{u}'_i\right)^2 $ 
			 \EndIf			
			 \State $\hat{s}_i=\sqrt{\hat{S}_i}\,\hat{x}_i$
			 \EndFor\label{forloop_messenger3}
   \State $s=\mathrm{ONT^{-1}}(\hat{s})$   
   \State \textbf{return} $s,\mat{\hat{S}}$ %\Comment{}
\EndProcedure
\end{algorithmic}
\end{algorithm}

\begin{figure*}
\centering{\includegraphics[width=1.0\textwidth,clip=true]{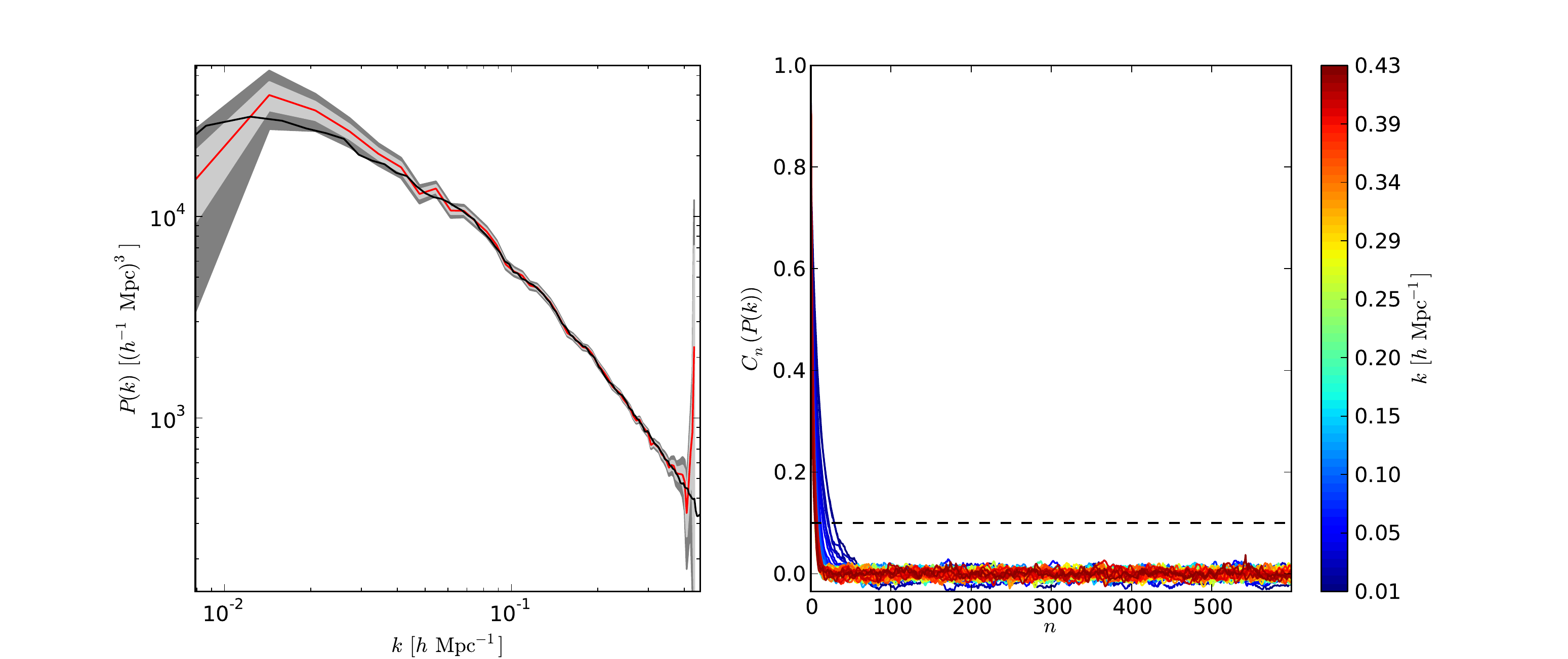}}
\caption{\label{fig:mean_corrlength} Ensemble mean power-spectrum (left panel) and correlation length (right panel) estimated from 80 000 samples. The left panel demonstrates the ensemble mean power spectrum (red curve) together with the corresponding one- (light grey) and two-sigma (dark grey) confidence regions. The black curve indicates the true underlying power-spectrum from which the mock signal was generated. As can be seen the inferred power-spectrum correctly follows the true underlying power-spectrum throughout the entire range of Fourier-modes. The right panel shows the correlation length between subsequent samples of power-spectrum amplitudes for different Fourier-modes as indicated by the colour coding. As can be seen, the correlation length quickly drops to zero indicating that independent power-spectrum samples can be generated at about every \(50\){\it{th}} step of the Markov Chain.   }

\end{figure*}

\subsection{Comments on parallelization}
\label{parellelization}
Generally Markov Chain Monte Carlo methods can be parallelized by either running several Markov chains in parallel or by distributing intra chain operations on multiple cores to speed up single serial chains. 
In particular, execution times for serial chain operation of our algorithm can be reduced by parallelizing loops over independent uni-variate sampling processes. In this case, care has to be taken when generating pseudo random numbers, a operation which is usually not thread-safe. A possible way, to prevent collisions between random numbers, is to assign individual pseudo random number generators with different seeds to each core respectively.
In this fashion, each process will independently generate a unique sequence of pseudo random numbers. 
Intra chain parallelization nevertheless still requires communication between different cores when performing ortho normal transforms, introducing waiting times for the fastest CPUs in the system.  
This parallelization scheme therefore is only reasonable when Markov chains exhibit long burn-in times and running parallel chains would be too wasteful. 
On the other hand, as will be demonstrated below, our algorithm shows short burn-in times and correlation lengths.
For this reason, running several completely independent Markov chains from over-dispersed initial conditions and with different respective seeds is favourable over intra chain parallelization.
As communication between independent Markov chains is not required, this approach will also show ideal scaling with the number of cores and independent samples, as generated by the sampling Process.

\section{Generation of a mock galaxy catalogue}
\label{mock_Data}
To demonstrate the numerical performance of the proposed method in a realistic setting, we will generate
artificial galaxy observations, following the approach previously described in \citet{JASCHE2010PSPEC} and \citet{JASCHE_SPEC2013}. In particular, we will perform a mock analysis on a cubic equidistant grid of side length \(1600 \,h^{-1}\mathrm{Mpc}\) consisting of \(128^3\) grid nodes.
The underlying cosmic density contrast field \(\delta_i\), being the signal to infer, is generated from a zero-mean normal distribution with the covariance matrix corresponding to a cosmological power spectrum, including baryon acoustic oscillations, generated via the prescription of \citet{1998ApJ...496..605E} and \citet{1999ApJ...511....5E}. Specifically, we  assume a standard \(\Lambda\)CDM cosmology  with the set of parameters given as  (\(\Omega_m=0.24\), \(\Omega_{\Lambda}=0.76\), \(\Omega_{b}=0.04\), \(h=0.73\), \(\sigma_8=0.74\), \(n_s=1\)).

As a next step, according to the likelihood described in Equation (\ref{eq:target_wiener_post}), this density field will be masked with the survey geometry and selection functions and normal distributed noise will be added.
Following the description in \citet{JASCHE_SPEC2013}, we aim to emulate characteristic features of the Sloan Digital Sky survey data release 7 (SDSS DR7) \citep[][]{SDSS7}.
In particular, we employ the redshift completeness of the SDSS DR7, which was computed with the MANGLE code provided by \citet{SWANSON2008MNRAS} and has been stored on a HEALPIX map with \(n_{side}=4096\) \citep[][]{HEALPIX}. Further, we assume a radial selection function following from a standard Schechter luminosity function with standard r-band parameters ( \(\alpha = -1.05\), \(M_* -5 \mathrm{log}_{10}(h)=-20.44\) ), and we limit the survey to only include galaxies within an apparent Petrosian r-band magnitude range  \(12.5\,<\,r<\,19.5\) and within the absolute magnitude ranges \(M_{min}=-21.3\) to \(M_{max}=-23.1\).
As usual, the radial selection function \(f(z)\) is then given by the integral of the Schechter luminosity function over the range in absolute magnitude.
The product of the two dimensional survey geometry \(M(\alpha_i,\delta_i)\) and the selection function \(f(z)\) at each point in the three dimensional volume yields the survey response operator:
\begin{equation}
R_i = M(\alpha_i,\delta_i) f^l(z_i) \, ,
\end{equation}
where \(\alpha_i\) and \(\delta_i\) are the right ascension and declination coordinates corresponding to the \(i\)th volume element, and \(z_i\) is the corresponding redshift.
Given these definitions and a realisation of the three dimensional density field \(\delta_i\), a realisation for the artificial galaxy number counts is obtained by:
\begin{equation}
N_i= \bar{N}\,R_i\,(1+\delta_i) + \sqrt{\bar{N}\,R_i}\, \epsilon_i \, ,
\end{equation}
where \(\epsilon_i\) is a white-noise field drawn from zero-mean and unit variance normal distribution, and the expected average number of galaxies \(\bar{N}\) is obtained via integration of the Schechter luminosity function by :
\begin{equation}
\bar{N} = \int_{M_{min}}^{M_{max}} \Phi(M)\, \mathrm{d}M \, . 
\end{equation}

\section{Results}
\label{Results}
In this section we discuss the results of the application to an artificial galaxy catalogue, with particular emphasis of the statistical efficiency of the algorithm. 

\subsection{Statistical efficiency}
\label{stat_eff}
Generally, in a Bayesian context the ill-posed inverse problem of inferring signals from observations, being subject to statistical uncertainty, is addressed by providing numerical representations of the the corresponding posterior
distribution. Here we achieve this goal via the proposed Gibbs sampling process, providing random realisations of the large scale structure and corresponding power-spectra conditioned to observations. This sampled representation of the posterior distribution then permits to address the inverse problem by providing summary statistics, accurately accounting for all uncertainties involved in the inference process. Nevertheless, Markov chain Monte Carlo methods generally draw random variates from the posterior distribution by generating a sequence of solutions satisfying ergodicity.

This approach generally yields a highly correlated sequence of solutions, which will almost surely converge to the target posterior distribution in the large sample limit.
As we seek to provide summary statistics for the parameters of interest, the performance of the proposed Gibbs sampler is determined by its ability to generate independent samples. This statistical efficiency
consequently is of crucial importance for any MCMC method, and is usually quantified by
the number of iterations required to generate a statistically independent sample. 
In order to determine the statistical efficiency of the proposed method, we will follow the standard procedure of estimating the correlation length of power-spectrum amplitudes within the chain \citep[see e.g.][]{2004ApJS..155..227E, JASCHE2010PSPEC, JASCHE_SPEC2013}.
Assuming all parameters in the Markov chain to be independent of each other, the correlation between subsequent power-spectrum amplitudes \(P(k)^i\) can be quantified in terms of the  auto-correlation function:
\begin{equation}
\label{eq:CORR_COEFF}
C(P(k))_n =\left \langle  \frac{P(k)^i-\left \langle P(k)\right \rangle}{\sqrt{\mathrm{Var} \left(P(k)\right)}} \frac{P(k)^{i+n}-\left \langle P(k)\right \rangle}{\sqrt{\mathrm{Var} \left(P(k)\right)}} \right \rangle \, ,
\end{equation}
where \(n\) is the distance in the chain measured in iterations. 
The results of this analysis are presented in the right panel of Figure \ref{fig:mean_corrlength}, where we estimated the correlation length from 80,000 recorded samples obtained by the application of our method to mock data.
In our tests we recorded every tenth sample generated by the Markov Chain.
Typically we determine the correlation length by the lag \(n_c\) in samples that is required for the auto-correlation function to drop below ten percent \citep[][]{2004ApJS..155..227E}.
Given this definition of correlation length, the test clearly indicates correlation lengths \(n_c\le50\) samples for all recorded power-spectrum modes.
Consequently, besides the ease of implementation the proposed method also exhibits excellent statistical efficiency for larges scale statistical applications
in modern cosmology and astrophysics.

\begin{figure*}
\centering{\includegraphics[width=1.0\textwidth,clip=true]{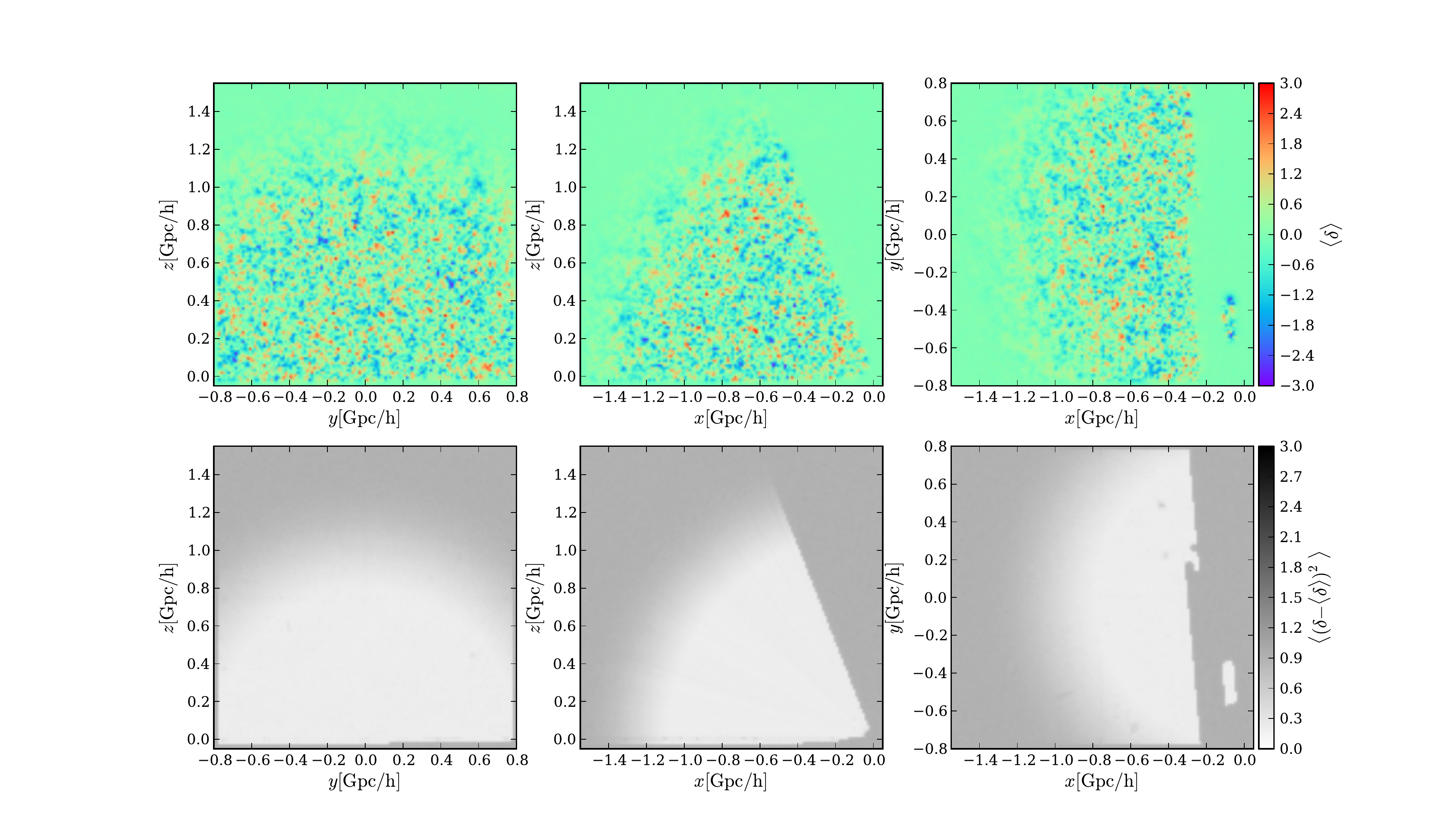}}
\caption{\label{fig:mean_var_dens} Marginalised posterior mean density (top panels) and variance (bottom panels) fields. The fields are estimated by taking the average over 80~000 samples of the posterior. The structure of those fields reflects the non-trivial survey geometry such as the one covered by the SDSS-DR7 main galaxy sample. We show a slice through $x=-750\, h^{-1}\rm{Mpc}$ (left panels), $y=0\, h^{-1}\rm{Mpc}$ (middle panels), $z=750\, h^{-1}\rm{Mpc}$ (right panels). }
\end{figure*}

\subsection{Inferring density maps and power-spectra}
\label{inf_spec_dens}
The messenger sampler, as proposed in this work, aims at the joint inference of a signal and its corresponding unknown correlation matrix.
Specifically, here, we propose to use this method for the joint inference of the cosmological three dimensional density field and its cosmological power-spectrum from observations.
As discussed above, inference of signals from noisy data is generally an ill-posed task, as there exists no unique solution.
The proposed method addresses this issue by exploring the joint posterior of the cosmic density field and the power-spectrum via an efficient
Gibbs sampling approach, providing a set of solutions being compatible with the observations. As a scientific result, this method therefore provides a numerical representation of the target posterior in terms
of random realisations of density fields and power-spectra conditioned to the observations. In this fashion, the ensemble of generated 
density fields and power-spectra permits us to generate any desired statistical summary of the parameters under consideration and to account for 
all joint and correlated uncertainties \citep[see also][]{2004ApJS..155..227E, JASCHE2010PSPEC, JASCHE_SPEC2013}.
As an example, we inferred the ensemble mean of the cosmic power-spectrum and corresponding uncertainties from the set of 80,000 generated power-spectrum samples.
The results for the ensemble mean power-spectrum together with the one and two sigma confidence regions are presented in left panel of Figure \ref{fig:mean_corrlength}.
It can be seen, that the inferred power-spectrum nicely follows the underlying fiducial power-spectrum from which the mock realisation was drawn.
Also, we do not observe any particular bias throughout the entire range of recovered Fourier modes. Note, as a remark, that these results
are generally compatible with previous approaches relying on different sampling strategies \citep[see e.g.][]{JASCHE2010PSPEC, JASCHE_SPEC2013}. 

Besides the cosmological power-spectrum, the method also provides maps of the three dimensional matter distribution.
In particular, in Figure \ref{fig:mean_var_dens} we demonstrate the ensemble mean density field estimated from 80,000 density samples along with the corresponding standard deviations, quantifying the uncertainty.
As anticipated from standard Wiener filtering approaches, the inferred density field recovers the underlying signal best in regions of high signal to noise and approaches mean density in regions of low signal to noise \citep[see e.g.][]{KITAURA2008,KITAURA2009,JASCHE2010PSPEC,JASCHE_SPEC2013}. The corresponding standard deviations for each volume element of the inferred density field are presented in the lower panels of Figure \ref{fig:mean_var_dens}  indicating corresponding uncertainties at all spatial points
in the observed domain. In this fashion the proposed method not only provides single estimates of the parameters under consideration but also provides thorough uncertainty quantification and means for error propagation.
These results therefore permit to derive any desired statistical summary and corresponding uncertainties, which are generally of crucial importance in order not to misinterpret the data.

\section{Summary and Conclusion}
\label{conclusion}
Modern cosmology has an ever increasing demand for fast and accurate statistical inference methods to counter present and upcoming avalanches of cosmological and astrophysical data. As pointed out in the introduction, inference of signals from observations subject to noise is a ill-posed problem requiring sophisticated statistical methods to quantify corresponding statistical uncertainties. Specifically large scale Bayesian inference, such as the joint inference of three dimensional matter density fields and corresponding cosmological power-spectra, from observations relies on complex and numerically expensive MCMC methods, often involving implementations of Krylov space methods or gradient based Hybrid Monte Carlo approaches \citep[see e.g.][]{KITAURA2008,JASCHE_SPEC2013}.    
Not only are these methods numerically expensive but are also hard to test and debug especially in large scale applications \citep[][]{COOK2006}. Besides these issues, the requirement to implement Krylov space or HMC methods constitutes a significant hurdle for rapid prototyping and development of large scale Bayesian inference methods in cosmology and astrophysics. 
To address these issues, in this work we present a new, efficient and trivial to implement Gibbs sampling approach for the joint inference of cosmological density fields and power-spectra for linear data models. This approach picks up basic ideas of the recently proposed messenger method for Wiener filtering, described by \citet{ELSNER2013},
and does not require any matrix inversions to explore high dimensional parameter spaces.  
As described in section \ref{augmented_posterior}, introducing a messenger field to mediate between different preferred bases, in which signal and noise covariance matrices can be expressed conveniently, yields an augmented Wiener posterior distribution which can be efficiently explored via a multiple block Gibbs sampling approach \citep{ELSNER2013}.
In particular, the proposed method turns the cumbersome approach of inverting multi-million dimensional matrices  into the task of sequentially drawing random numbers from only uni-variate normal distributions. While trivial to implement, iteration of this process yields full multi-variate random fields drawn from the desired target Wiener posterior distribution, hence correctly addressing the large scale inference problem.
To address also problems in which the covariance matrix of the signal to infer is unknown, we add a power-spectrum block sampler to jointly infer the signal and it's power spectrum. 

As described in section \ref{implementation}, this power-spectrum sampler permits to efficiently explore the high as well as the low signal to noise regime. While low signal to noise regimes are dominated by stochastic noise, high signal to noise regions are dominated solely by the much smaller sample variance. Using this sample variance to determine step sizes in Markov transitions will result in correct sampling of the target posterior, but may also yield unfeasibly long Markov chains as parameters in the low signal-to-noise regime remain correlated over many steps. To counter this problem, in \ref{implementation} we introduced a coordinate transform which permits to perform larger steps in low-signal-to noise regimes via a Metropolis Hastings transition step. The combination of both approaches yields a covariance matrix sampler that is efficient at all regimes, while only requiring the ability to
generate uni-variate random numbers. 

In this fashion the task of jointly sampling a signal and its covariance matrix can be addressed purely by a sequence of uni-variate sampling processes. 

In section \ref{Results} we exemplify the performance of our method in a cosmological setting by applying it to a artificial galaxy mock catalogue, described previously in section \ref{mock_Data}, aiming at the joint inference of the three dimensional density distribution and its cosmological power-spectrum from observations. This artificial data set emulates dominant features of the Sloan Digital Sky Survey data release 7, in particular survey geometry, selection effects and noise, and thus constitutes a realistic test scenario. 

A particular important aspect, when dealing with Markov Chain Monte Carlo algorithms, is the determination of their statistical efficiency. As any Markov Chain Monte Carlo method generates a sequence of correlated samples the amount of actually produced independent samples is limited by the total length of the chain. In section \ref{stat_eff} we therefore analysed the intra-chain correlation length between subsequently generated samples of the cosmological power-spectrum. These test demonstrates formidable statistical efficiency for the proposed method
over the entire range of Fourier-modes present in the analysis. Specifically these tests indicate that the proposed Markov algorithm generates independent samples at every 50{\it{th}} iteration of the Markov chain, where we chose one cycle to consists in ten Markov transitions.

Section \ref{inf_spec_dens} discusses the results obtained by the proposed Markov method. In particular, the method provides estimates for the ensemble mean cosmological power-spectrum and corresponding uncertainty quantification. The inferred power-spectrum recovers the underlying true signal and shows no sign of bias throughout the entire range of Fourier-modes under consideration. These result are also consistent with previous results \citep[see ][]{JASCHE2010PSPEC,JASCHE_SPEC2013}.  

Furthermore, in our example case, the method also provides inferred three dimensional maps of the cosmic matter distribution. In Figure \ref{fig:mean_var_dens} we demonstrate ensemble mean estimates of the density field and ensemble covariance maps quantifying the corresponding uncertainty. The proposed method therefore not only provides single estimates of signals, but also provides means to quantify and propagate statistical uncertainties for any finally inferred quantity, as is required for modern precision cosmology.

The ease of implementation, numerical and statistical efficiency renders this method an ideal tool for large scale Bayesian applications involving million dimensional problems and linear data models. A particularly important feature of the method is, that it only requires the ability to sample from uni-variate distributions and thus can be trivially implemented and tested by even inexperienced users or can be used for rapid prototyping and development of more complex inference frameworks.

In summary, we propose a statistical and numerically efficient Gibbs sampling approach for the inference of a unknown signal and its covariance matrix from observations subject to statistical uncertainties and systematics. Particularly due to the ease of implementation we anticipate this method to greatly add to the propagation of high precision large scale data analysis methods in cosmology and astrophysics, eventually leading to a more complete understanding of our Universe.

\section*{Acknowledgements}
We thank Benjamin Wandelt and Franz Elsner for very useful discussions and comments. 
Special thanks also go to St\'ephane Rouberol for his support during the course of this work, in particular for guaranteeing flawless use of all required computational resources.
JJ is partially supported by a Feodor Lynen Fellowship by the Alexander von Humboldt foundation and Benjamin Wandelt's Chaire d'Excellence from the Agence Nationale de la Recherche.
This work made in the ILP LABEX (under reference ANR-10-LABX-63) was supported by French state funds
managed by the ANR within the Investissements dAvenir programme under reference ANR-11-IDEX-0004-02.

\bibliographystyle{mn2e}

\appendix

%\bsp

\label{lastpage}

\end{document}